\documentclass{article}

\PassOptionsToPackage{numbers, sort&compress}{natbib}

\usepackage[dblblindworkshop, final]{neurips_2026}

\usepackage[utf8]{inputenc} 
\usepackage[T1]{fontenc}    
\usepackage[hidelinks]{hyperref} 
\usepackage{url}            
\usepackage{booktabs}       
\usepackage{tabularx}       
\usepackage{amsfonts}       
\usepackage{nicefrac}       
\usepackage{microtype}      
\usepackage{xcolor}         
\usepackage{graphicx}       

\usepackage[acronym, nomain, nopostdot]{glossaries}

\newacronym{om}{OM}{optical module}
\newacronym{ml}{ML}{machine learning}
\newacronym{mc}{MC}{Monte Carlo}
\newacronym{bce}{BCE}{binary cross-entropy}
\newacronym{mse}{MSE}{mean squared error}
\newacronym{dl}{DL}{deep learning}
\newacronym{mlp}{MLP}{multi-layer perceptron}
\newacronym{vmf}{vMF}{von Mises--Fisher}
\newacronym{ssl}{SSL}{self-supervised learning}
\newacronym{nsf}{NSF}{National Science Foundation}

\workshoptitle{Representations for the Physical Sciences Workshop}
\title{CURTAIN: Towards a Foundation Model for Neutrino Telescopes}

\author{
  Severin Magel \\
  Harvard University \\
  Technical University of Munich \\
  \texttt{severin.magel@tum.de} \\
  \And
  Felix J. Yu \\
  Harvard University \\
  \texttt{felixyu@g.harvard.edu} \\
  \And
  Carlos A. Argüelles \\
  Harvard University \\
  \texttt{carguelles@fas.harvard.edu} \\
}

\begin{document}

\maketitle

\begin{abstract}
  Neutrino telescopes record more than a trillion photon detections per year, yet there does not exist a \gls{ssl} task which fully satisfies all three conditions: improving as the pretraining set grows, transferring to a different detector, and outperforming the best model trained without pretraining at all. We introduce \textbf{\texttt{CURTAIN}}, a temporal masking scheme and the first \gls{ssl} task for neutrino telescopes that achieves all of these goals. Downstream performance monotonically improves with the number of pretraining events, as predicted by scaling laws. Our encoder builds on \texttt{DeepIce}, currently the leading model on the NuBench benchmarks direction reconstruction task. Pretraining it with \texttt{CURTAIN} reduces its median angular error by $45\,\%$ and its median energy error by $31\,\%$ compared to training from scratch. It also transfers to a detector geometry never seen in pretraining, reducing the median angular error by $36\,\%$ compared to training from scratch. At matched capacity and pretraining data, \texttt{CURTAIN}-pretrained encoders beat those pretrained by existing \gls{ssl} tasks on both tasks and at every pretraining set size.
\end{abstract}

\glsresetall

\section{Introduction}

\begin{figure}[t]
  \centering
  \includegraphics[width=\textwidth]{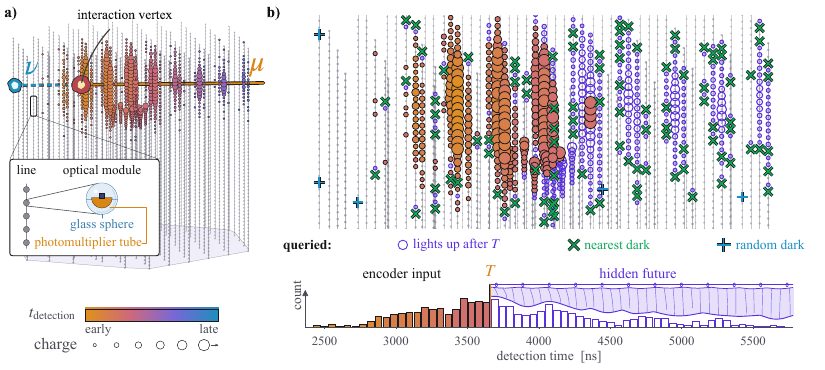}
  \caption{
    \textbf{(a)} A neutrino telescope, here on IceCube's array geometry: a neutrino ($\nu$) interacts and the outgoing muon ($\mu$) lights \glsentryshortpl{om} along its path. Marker color is arrival time, area is charge.
    \textbf{(b)} \texttt{CURTAIN}'s masking scheme on the same simulated event, cut at the threshold $T$. Filled circles are the encoder's input. The model is queried at three kinds of positions: open circles are \glsentryshortpl{om} that light up only after $T$ (positives); green crosses are the dark \glsentryshortpl{om} nearest to each positive (hard negatives); blue plus signs are dark \glsentryshortpl{om} drawn at random (easy negatives). Below, the hit-time distribution, curtained off after $T$.
  }
  \label{fig:curtain_overview}
\end{figure}

Neutrinos interact with matter so rarely that they reach us in a straight line from their sources, carrying a direct view of the most violent processes in the universe~\citep{spurioProbesMultimessengerAstrophysics2018}. That same rarity makes them hard to detect. A neutrino telescope instruments a km$^{3}$-scale volume of glacial ice or deep water with vertical lines of \glspl{om} (Figure~\ref{fig:curtain_overview}(a)). A neutrino interacting in the medium creates charged particles, which emit Cherenkov light as they travel. The \glspl{om} detect that light, registering the interaction as a variable-length spatio-temporal point cloud called an event. Each of its points, a pulse, records one or more detected photons and carries a position, an arrival time, and the amount of light detected, which commonly is referred to as charge.

Every physics analysis must estimate the neutrino's direction and energy from this
light pattern. Real data carry no labels, so estimators rely on high-fidelity \gls{mc}
simulation: \gls{ml} models are trained on pattern--target pairs from this
forward model and applied to real detected
patterns~\citep{abbasiConvolutionalNeuralNetwork2021}. This simulation-based inference setup carries two costs. Every
downstream task requires large simulation datasets, and generating them is
expensive: the Cherenkov light of every event must be propagated photon by photon
through hundreds of meters of ice or water. Any mismatch between the simulated and the
real data distribution also biases the estimator in ways that are difficult to
quantify~\citep{aartsenEfficientPropagationSystematic2019}. Such mismatch is hard
to avoid, because the detecting medium is a natural one that was never
characterized in a laboratory. The optical properties of the glacial ice must be
calibrated in situ and vary strongly with depth~\citep{aartsenIceCubeNeutrinoObservatory2017}, while
water detectors sit in a time-varying optical background of bioluminescence and
radioactive decay~\citep{adrian-martinezLetterIntentKM3NeT2016}.

\Gls{ssl} offers a way around both costs. Foundation models
pretrained on large unlabeled datasets have transformed many fields, among them language, vision, and
robotics~\citep{brownLanguageModelsAre2020a, MAEVision, nvidiaGR00TN1Open2025}. Across the operating detectors, neutrino telescopes record more than a trillion pulses per year. This is a very conservative lower bound (Appendix~\ref{app:datavolume}), and it already exceeds the 300 billion tokens GPT-3 was trained on~\citep{brownLanguageModelsAre2020a}. A shared, information-rich representation of these patterns would improve every downstream analysis at once, and cut the number of simulated particle interactions each new analysis has to generate. Learning representations of events from real data can also improve the estimator's generalization from \gls{mc} to data.

\Gls{ssl} has already been applied to neutrino telescopes. \texttt{PolarBERT}~\citep{timiryasovPolarBERTFoundationModela} improves direction estimation from pretraining, reducing the reliance on labeled simulation, and \texttt{Neptune}~\citep{yuReducingSimulationDependence2026a} uses pretraining to mitigate \gls{mc} mismodeling. We argue that neither is yet a foundation model for neutrino telescopes, which we take to require three properties: it improves as the pretraining set grows, it generalizes beyond a single detector, and it outperforms the best model trained without pretraining. \texttt{PolarBERT}'s input vocabulary enumerates the sensors of one detector, and neither method can pretrain the strongest supervised architecture. In our matched-capacity comparison, \texttt{Neptune}'s downstream performance does not improve with more pretraining data, and \texttt{PolarBERT} stays far behind a model trained from scratch on direction.

We introduce \textbf{\texttt{CURTAIN}}, an \gls{ssl} pretraining method for neutrino telescope data that masks events in time rather than pulse by pulse. Its encoder--decoder architecture does not require passing masked pulses through the encoder, allowing the use of time- and position-dependent attention mechanisms. It is the first \gls{ssl} task to satisfy all three properties.
\section{\texttt{CURTAIN}}
\label{sec:curtain}

Like \texttt{PolarBERT} and \texttt{Neptune}, \texttt{CURTAIN} follows the masked modeling paradigm~\citep{MAEVision, pangMaskedAutoencodersPoint2022}: hide part of an event and ask the model to recover it from the rest. We split each event at a randomly sampled threshold time $T$ (Figure~\ref{fig:curtain_overview}(b)): pulses before $T$ are the visible context, and every \gls{om} lit only after $T$ is held out with all of its information hidden. The model is then queried at sensor positions outside the context, each query carrying nothing but a position, and must predict whether that sensor lights up after $T$. The queried positions are of two kinds: the held-out \glspl{om}, which do light up after $T$ (positives), and \glspl{om} that stay dark for the whole event (negatives), taken either nearest to a positive (hard) or at random (easy). An additional auxiliary head regresses the arrival time. Appendix~\ref{app:architecture} gives the architecture and training details.

This masking scheme is in contrast to \texttt{PolarBERT}, which hides a random subset of pulses~\citep{timiryasovPolarBERTFoundationModela}, and \texttt{Neptune}, which in the variant we compare against hides the arrival time of every pulse~\citep{yuReducingSimulationDependence2026a}; Appendix~\ref{app:schemes} compares the three schemes. We conjecture that masking temporally is what makes a representation transfer. A randomly masked pulse can be recovered by interpolating between its neighbors in space and time; masking everything after $T$ removes that shortcut and forces the model to learn how light propagates through the detector.

\begin{figure}[t]
  \centering
  \includegraphics[width=\textwidth]{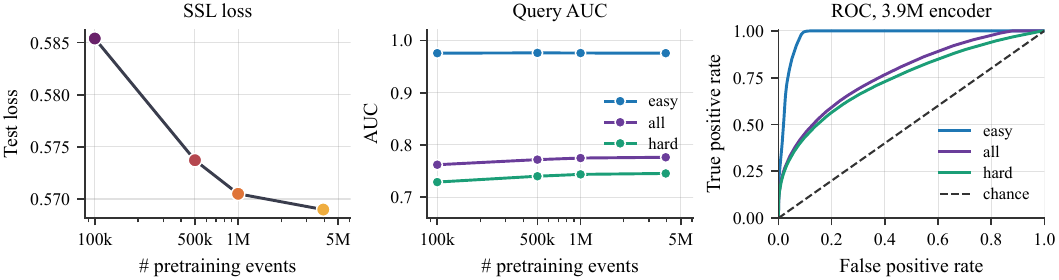}
  \caption{
    \texttt{CURTAIN} pretraining performance. \textbf{Left}: held-out \glsentryshort{ssl} loss against the number of pretraining events. \textbf{Middle}: query ROC-AUC for separating the positives from the easy negatives, from the hard negatives, and from all negatives together. \textbf{Right}: ROC curves of the encoder pretrained on 3.9 million events.
  }
  \label{fig:pretraining-scaling}
\end{figure}

This design is also why \texttt{CURTAIN}, unlike \texttt{Neptune} and \texttt{PolarBERT}, can pretrain the strongest supervised architecture. Both baselines are encoder-only in pretraining, so masked pulses must pass through the encoder's attention blocks. \texttt{DeepIce} (second-place solution in \citet{bukhariIceCubeNeutrinosDeep2023}), the leading model on the direction reconstruction task of the NuBench benchmark~\citep{orsoeNuBenchOpenBenchmark2026a}, uses \texttt{SpaceTime} attention: a bias derived from the space--time distance between two pulses, undefined when a position or time is masked. \texttt{CURTAIN}'s encoder sees only visible pulses and its decoder is queried at the held-out positions, so any encoder whose attention bias depends on position or time can be pretrained this way, \texttt{DeepIce} included.

\section{Experimental setup}

All data used in this work comes from the NuBench datasets~\citep{orsoeNuBenchOpenBenchmark2026a}. We use only the \texttt{hexagon} and \texttt{flower\_l} datasets, which resemble the IceCube~\citep{aartsenIceCubeNeutrinoObservatory2017} and KM3NeT-ARCA~\citep{adrian-martinezLetterIntentKM3NeT2016} geometries with simplified detector electronics, optical modules and optical medium, and restrict ourselves to track events, in which the incident neutrino creates a muon that travels through the detector.

We pretrain a \texttt{DeepIce} encoder using the \texttt{CURTAIN} \gls{ssl} task. The encoder is downscaled to 2.2 million parameters to keep the computational cost manageable. We pretrain on 100 thousand, 500 thousand, 1 million, and 3.9 million events of the \texttt{hexagon} dataset using early stopping. We use a \gls{bce} loss on whether each queried sensor lights up after $T$, and a Huber loss on the first-hit time of the sensors that do. As baselines we pretrain \texttt{Neptune} and \texttt{PolarBERT} on the same data, at the same depth, model dimension and number of heads, so all schemes are compared at matched capacity.

We finetune the pretrained encoders on direction and energy reconstruction, on a training set of 100 thousand \texttt{hexagon} events. This split keeps only events with at least eight pulses, shares no event with any pretraining set, and has 200 thousand validation and 200 thousand test events. We use the \gls{vmf} negative log likelihood for direction and the \gls{mse} on $\log_{10} E$ for energy. Both heads are parameterized differently from \citet{orsoeNuBenchOpenBenchmark2026a}, which we found to train more stably (Appendix~\ref{app:finetuning}). 

We additionally finetune the direction task on 100 thousand \texttt{flower\_l} events, with 20 thousand validation and 50 thousand test events, to test how well the pretrained representations generalize to an unseen geometry. All performance is evaluated on held-out sets drawn from the same distribution as the corresponding training set and disjoint from the pretraining events, reporting the median angular error, the angle between the reconstructed and the true direction, and the median $|\Delta \log_{10} E|$, evaluated in log space given the large dynamic range of neutrino events~\citep{spurioProbesMultimessengerAstrophysics2018}. Each reported value comes from a single training run, so we give no estimate of run-to-run variation.

\section{Results}

\subsection{Pretraining performance}

Figure \ref{fig:pretraining-scaling} shows the performance of \texttt{CURTAIN} in pretraining. The held-out \gls{ssl} loss falls monotonically with pretraining set size, from 0.585 at 100 thousand events to 0.569 at 3.9 million (left panel in Figure~\ref{fig:pretraining-scaling}), consistent with data scaling at fixed model size~\citep{kaplanScalingLawsNeural2020}. The easy negatives are already almost perfectly separated from the positives by the encoder pretrained on the smallest set (AUC $= 0.976$), and this separation does not improve with pretraining set size. The hard-negative AUC instead rises from 0.729 to 0.745, carrying the overall query AUC from 0.762 to 0.776.

\begin{figure}[t]
  \centering
  \includegraphics[width=\textwidth]{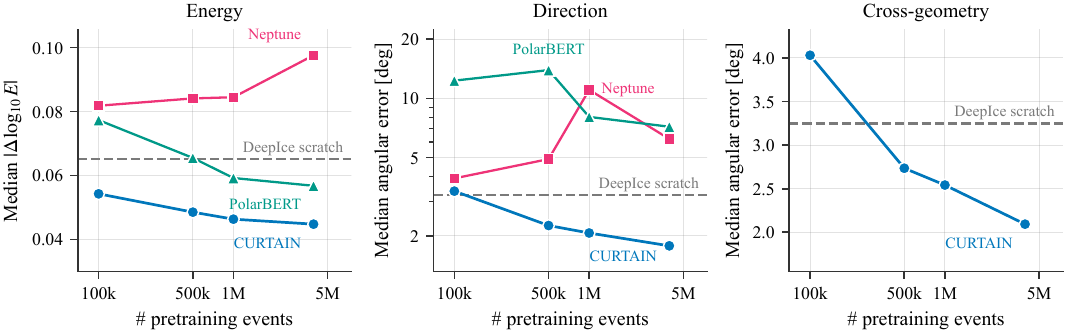}
  \caption{
    Downstream performance against the number of pretraining events, after finetuning on 100 thousand labeled events. \textbf{Left} and
    \textbf{middle}: median energy and angular error on \texttt{hexagon}, the geometry also used for pretraining, for \texttt{CURTAIN} and the \texttt{Neptune} and \texttt{PolarBERT} baselines; the angular-error axis is logarithmic. \textbf{Right}: median angular error on \texttt{flower\_l}, a geometry never seen during pretraining, for \texttt{CURTAIN} only. Dashed lines: \texttt{DeepIce} trained from scratch on the same labeled events.
  }
  \label{fig:finetune_scaling}
\end{figure}

\subsection{Finetuning tasks}

Direction reconstruction improves monotonically with pretraining set size
(Figure~\ref{fig:finetune_scaling}, middle), reaching $1.78^\circ$ against
$3.22^\circ$ from scratch, a $45\,\%$ reduction. Only the encoder pretrained on the smallest set, 100 thousand events, is slightly worse than training from scratch ($3.37^\circ$). Energy also improves monotonically with pretraining set size (Figure~\ref{fig:finetune_scaling}, left), and unlike direction it beats training from scratch already at 100 thousand events (0.0543). It reaches a median
$|\Delta \log_{10} E|$ of 0.0448 against 0.0651, a $31\,\%$ reduction and the smaller gain of the two.
At matched capacity, pretraining data and compute budget, \texttt{CURTAIN} beats both
baselines on both tasks at every pretraining set size.

This comparison is between complete methods: each baseline brings its own encoder and input embedding along with its pretraining objective, so it does not isolate the contribution of temporal masking. The comparison against the from-scratch \texttt{DeepIce} is the one that holds the architecture fixed. \texttt{Neptune}'s downstream performance does not improve with the pretraining
set size. \texttt{PolarBERT} stays far behind the from-scratch \texttt{DeepIce} on direction at every size we tried ($7.17^\circ$ at best), and on energy only draws level with it at 500 thousand events and overtakes it from 1 million. Its weak direction performance is plausibly because \texttt{PolarBERT}'s \gls{om}-token embedding carries no geometric prior, so it needs large amounts of data to become competitive.

\subsection{Cross-detector generalization}

An encoder pretrained on \texttt{hexagon} also transfers to \texttt{flower\_l}, a geometry it never saw during pretraining (Figure~\ref{fig:finetune_scaling}, right). With 100 thousand labeled events it improves on training from scratch, from $3.25^\circ$ to $2.09^\circ$ ($36\,\%$), using the encoder pretrained on 3.9 million events. A \texttt{CURTAIN} query carries only a sensor position, so nothing the encoder learns is tied to a sensor numbering, which makes transfer to another layout possible; whether temporal masking is what makes the transferred representation useful, as conjectured in Section~\ref{sec:curtain}, would require further ablation studies. Transfer also improves monotonically with pretraining set size, from $4.03^\circ$ at 100 thousand events, still worse than training from scratch, to $2.09^\circ$ at 3.9 million; pretraining overtakes training from scratch from 500 thousand events ($2.73^\circ$).

\section{Conclusion}

We introduced \texttt{CURTAIN}, an \gls{ssl} pretraining task for neutrino telescopes that temporally splits an event at a randomly sampled threshold time and requires the model to predict which sensors light up after that time. We asked three things of a foundation model, and \texttt{CURTAIN} delivers all three at the scale we tested. Its performance improves monotonically with the size of the pretraining set, in the pretext task and in both downstream tasks. The representations generalize beyond a single detector, reaching a median angular error of $2.09^\circ$ against $3.25^\circ$ from scratch on a geometry never seen during pretraining, given 100 thousand labeled events. And it outperforms the best model trained without pretraining, cutting the median angular error by $45\,\%$ and the median $|\Delta \log_{10} E|$ by $31\,\%$. At matched capacity and pretraining budget, \texttt{CURTAIN} also beats both \texttt{Neptune} and \texttt{PolarBERT} on both downstream tasks by a large margin. All of this was shown in simulation, with simplified detector electronics, optical modules and optical medium, from single training runs, and after full finetuning on 100 thousand labeled events; real data, label efficiency, and an ablation that separates temporal masking from the encoder remain open. \texttt{CURTAIN} was pretrained on 3.9 million simulated events, less than $10^{-4}$ of what IceCube records in a year. Pretraining at that scale, and on real rather than simulated data, is the natural next step. In the future we plan on jointly training on different detectors and exploring model-size and compute-constrained scaling laws of the \gls{ssl} and finetuning tasks.

%
\begin{ack}

SM is supported by a PROMOS scholarship from the German Academic Exchange Service (Deutscher Akademischer Austauschdienst, DAAD), awarded by the Technical University of Munich. FJY is supported by the NSF Graduate Research Fellowship under Grant No. 2140743. 
CAA are supported by the Faculty of Arts and Sciences of Harvard University, the National Science Foundation (NSF, CAREER Grant No. 2239795), the John Templeton Foundation (Grant No. 63651), the Research Corporation for Science Advancement, and the David \& Lucile Packard Foundation.
\end{ack}

\bibliographystyle{unsrtnat3}
\bibliography{references}

\begin{thebibliography}{18}
\providecommand{\natexlab}[1]{#1}
\providecommand{\url}[1]{\texttt{#1}}
\expandafter\ifx\csname urlstyle\endcsname\relax
  \providecommand{\doi}[1]{doi: #1}\else
  \providecommand{\doi}{doi: \begingroup \urlstyle{rm}\Url}\fi

\bibitem[Spurio(2018)]{spurioProbesMultimessengerAstrophysics2018}
Maurizio Spurio.
\newblock \emph{Probes of {{Multimessenger Astrophysics}}: {{Charged}} Cosmic Rays, Neutrinos, {$\gamma$}-Rays and Gravitational Waves}.
\newblock Astronomy and {{Astrophysics Library}}. Springer International Publishing, Cham, 2018.
\newblock ISBN 978-3-319-96853-7 978-3-319-96854-4.
\newblock \doi{10.1007/978-3-319-96854-4}.

\bibitem[Abbasi et~al.(2021)Abbasi, Ackermann, Adams, Aguilar, Ahlers, Ahrens, Alispach, Alves, Amin, An, Andeen, Anderson, Ansseau, Anton, Arg{\"u}elles, Axani, Bai, Balagopal~V., Barbano, Barwick, Bastian, Basu, Baum, Baur, Bay, Beatty, Becker, Becker~Tjus, Bellenghi, BenZvi, Berley, Bernardini, Besson, Binder, Bindig, Blaufuss, Blot, B{\"o}ser, Botner, B{\"o}ttcher, Bourbeau, Bourbeau, Bradascio, Braun, Bron, {Brostean-Kaiser}, Burgman, Busse, Campana, Chen, Chirkin, Choi, Clark, Clark, Classen, Coleman, Collin, Conrad, Coppin, Correa, Cowen, Cross, Dave, Clercq, DeLaunay, Dembinski, Deoskar, Ridder, Desai, Desiati, De~Vries, De~Wasseige, De~With, DeYoung, Dharani, Diaz, {D{\'i}az-V{\'e}lez}, Dujmovic, Dunkman, DuVernois, Dvorak, Ehrhardt, Eller, Engel, Evans, Evenson, Fahey, Fazely, Fiedlschuster, Fienberg, Filimonov, Finley, Fischer, Fox, Franckowiak, Friedman, Fritz, F{\"u}rst, Gaisser, Gallagher, Ganster, Garrappa, Gerhardt, Ghadimi, Glaser, Glauch, Gl{\"u}senkamp, Goldschmidt, Gonzalez, Goswami,
  Grant, Gr{\'e}goire, Griffith, Griswold, G{\"u}nd{\"u}z, Haack, Hallgren, Halliday, Halve, Halzen, Minh, Hanson, Hardin, Harnisch, Haungs, Hauser, Hebecker, Helbing, Henningsen, Hettinger, Hickford, Hignight, Hill, Hill, Hoffman, Hoffmann, Hoinka, {Hokanson-Fasig}, Hoshina, Huang, Huber, Huber, Hultqvist, H{\"u}nnefeld, Hussain, In, Iovine, Ishihara, Jansson, Japaridze, Jeong, Jones, Joppe, Kang, Kang, Kang, Kappes, Kappesser, Karg, Karl, Karle, Katz, Kauer, Kellermann, Kelley, Kheirandish, Kim, Kin, Kintscher, Kiryluk, Klein, Koirala, Kolanoski, K{\"o}pke, Kopper, Kopper, Koskinen, Koundal, Kovacevich, Kowalski, Krings, Kr{\"u}ckl, Kurahashi, Kyriacou, Lagunas~Gualda, Lanfranchi, Larson, Lauber, Lazar, Leonard, Leszczy{\'n}ska, Li, Liu, Lohfink, Lozano~Mariscal, Lu, Lucarelli, Ludwig, Luszczak, Lyu, Ma, Madsen, M.~Mahn, Makino, Mallik, Mancina, Mari{\c s}, Maruyama, Mase, McNally, Meagher, Medina, Meier, {Meighen-Berger}, Merz, Micallef, Mockler, Moment{\'e}, Montaruli, Moore, Morik, Morse, Moulai, Naab,
  Nagai, Naumann, Necker, Nguy{\~{\^e}}n, Niederhausen, Nisa, Nowicki, Nygren, Obertacke~Pollmann, Oehler, Olivas, O'Sullivan, Pandya, Pankova, Park, Parker, Paudel, Peiffer, P{\'e}rez De Los~Heros, Philippen, Pieloth, Pieper, Pizzuto, Plum, Popovych, Porcelli, Prado~Rodriguez, Price, Pries, Przybylski, Raab, Raissi, Rameez, Rawlins, Rea, Rehman, Reimann, Renschler, Renzi, Resconi, Reusch, Rhode, Richman, Riedel, Robertson, Roellinghoff, Rongen, Rott, Ruhe, Ryckbosch, Rysewyk~Cantu, Safa, Sanchez~Herrera, Sandrock, Sandroos, Santander, Sarkar, Sarkar, Satalecka, Scharf, Schaufel, Schieler, Schlunder, Schmidt, Schneider, Schneider, Schr{\"o}der, Schumacher, Sclafani, Seckel, Seunarine, Sharma, Shefali, Silva, Skrzypek, Smithers, Snihur, Soedingrekso, Soldin, Spiczak, Spiering, Stachurska, Stamatikos, Stanev, Stein, Stettner, Steuer, Stezelberger, Stokstad, St{\"u}rwald, Stuttard, Sullivan, Taboada, Tenholt, {Ter-Antonyan}, Tilav, Tischbein, Tollefson, Tomankova, T{\"o}nnis, Toscano, Tosi, Trettin, Tselengidou,
  Tung, Turcati, Turcotte, Turley, Twagirayezu, Ty, Unland~Elorrieta, {Valtonen-Mattila}, Vandenbroucke, Van~Eijk, Van~Eijndhoven, Vannerom, Van~Santen, Verpoest, Vraeghe, Walck, Wallace, Watson, Weaver, Weindl, Weiss, Weldert, Wendt, Werthebach, Weyrauch, Whelan, Whitehorn, Wiebe, Wiebusch, Williams, Wolf, Woschnagg, Wrede, Wulff, Xu, Xu, Yanez, Yoshida, Yuan, and Zhang]{abbasiConvolutionalNeuralNetwork2021}
R.~Abbasi, M.~Ackermann, J.~Adams, et~al.
\newblock A convolutional neural network based cascade reconstruction for the {{IceCube Neutrino Observatory}}.
\newblock \emph{Journal of Instrumentation}, 16\penalty0 (07):\penalty0 P07041, July 2021.
\newblock ISSN 1748-0221.
\newblock \doi{10.1088/1748-0221/16/07/P07041}.

\bibitem[Aartsen et~al.(2019)Aartsen, Ackermann, Adams, Aguilar, Ahlers, Alispach, Atoum, Andeen, Anderson, Ansseau, Anton, Arg{\"u}elles, Auffenberg, Axani, Backes, Bagherpour, Bai, V., Barbano, Barwick, Bastian, Baum, Baur, Bay, Beatty, Becker, Tjus, BenZvi, Berley, Bernardini, Besson, Binder, Bindig, Blaufuss, Blot, Bohm, B{\"o}rner, B{\"o}ser, Botner, B{\"o}ttcher, Bourbeau, Bourbeau, Bradascio, Braun, Bron, {Brostean-Kaiser}, Burgman, Buscher, Busse, Carver, Chen, Cheung, Chirkin, Choi, Clark, Classen, Coleman, Collin, Conrad, Coppin, Correa, Cowen, Cross, Dave, Clercq, DeLaunay, Dembinski, Deoskar, Ridder, Desiati, {de Vries}, {de Wasseige}, {de With}, DeYoung, Diaz, {D{\'i}az-V{\'e}lez}, Dujmovic, Dunkman, Dvorak, Eberhardt, Ehrhardt, Eller, Engel, Evenson, Fahey, Fazely, Felde, Filimonov, Finley, Franckowiak, Friedman, Fritz, Gaisser, Gallagher, Ganster, Garrappa, Gerhardt, Ghorbani, Glauch, Gl{\"u}senkamp, Goldschmidt, Gonzalez, Grant, Griffith, Griswold, G{\"u}nder, G{\"u}nd{\"u}z, Haack, Hallgren,
  Halve, Halzen, Hanson, Haungs, Hebecker, Heereman, Heix, Helbing, Hellauer, Henningsen, Hickford, Hignight, Hill, Hoffman, Hoffmann, Hoinka, {Hokanson-Fasig}, Hoshina, Huang, Huber, Huber, Hultqvist, H{\"u}nnefeld, Hussain, In, Iovine, Ishihara, Japaridze, Jeong, Jero, Jones, Jonske, Joppe, Kang, Kang, Kappes, Kappesser, Karg, Karl, Karle, Katz, Kauer, Kelley, Kheirandish, Kim, Kintscher, Kiryluk, Kittler, Klein, Koirala, Kolanoski, K{\"o}pke, Kopper, Kopper, Koskinen, Kowalski, Krings, Kr{\"u}ckl, Kulacz, Kurahashi, Kyriacou, Labare, Lanfranchi, Larson, Lauber, Lazar, Leonard, Leszczy{\'n}ska, Leuermann, Liu, Lohfink, Mariscal, Lu, Lucarelli, L{\"u}nemann, Luszczak, Lyu, Ma, Madsen, Maggi, Mahn, Makino, Mallik, Mallot, Mancina, Mari{\c s}, Maruyama, Mase, Maunu, McNally, Meagher, Medici, Medina, Meier, {Meighen-Berger}, Menne, Merino, Meures, Micallef, Mockler, Moment{\'e}, Montaruli, Moore, Morse, Moulai, Muth, Nagai, Naumann, Neer, Niederhausen, Nowicki, Nygren, Pollmann, Oehler, Olivas, O'Murchadha,
  O'Sullivan, Palczewski, Pandya, Pankova, Park, Peiffer, {P{\'e}rez de los Heros}, Philippen, Pieloth, Pinat, Pizzuto, Plum, Porcelli, Price, Przybylski, Raab, Raissi, Rameez, Rauch, Rawlins, Rea, Reimann, Relethford, Renschler, Renzi, Resconi, Rhode, Richman, Robertson, Rongen, Rott, Ruhe, Ryckbosch, Rysewyk, Safa, Herrera, Sandrock, Sandroos, Santander, Sarkar, Sarkar, Satalecka, Schaufel, Schieler, Schlunder, Schmidt, Schneider, Schneider, Schr{\"o}der, Schumacher, Sclafani, Seckel, Seunarine, Shefali, Silva, Snihur, Soedingrekso, Soldin, Song, Spiczak, Spiering, Stachurska, Stamatikos, Stanev, Stein, Steinm{\"u}ller, Stettner, Steuer, Stezelberger, Stokstad, St{\"o}{\ss}l, Strotjohann, St{\"u}rwald, Stuttard, Sullivan, Taboada, Tenholt, {Ter-Antonyan}, Terliuk, Tilav, Tollefson, Tomankova, T{\"o}nnis, Toscano, Tosi, Trettin, Tselengidou, Tung, Turcati, Turcotte, Turley, Ty, Unger, Elorrieta, Usner, Vandenbroucke, Driessche, van Eijk, van Eijndhoven, Vanheule, van Santen, Vraeghe, Walck, Wallace,
  Wallraff, Wandkowsky, Watson, Weaver, Weindl, Weiss, Weldert, Wendt, Werthebach, Whelan, Whitehorn, Wiebe, Wiebusch, Wille, Williams, Wills, Wolf, Wood, Wood, Woschnagg, Wrede, Xu, Xu, Xu, Yanez, Yodh, Yoshida, Yuan, and Z{\"o}cklein]{aartsenEfficientPropagationSystematic2019}
M.G. Aartsen, M.~Ackermann, J.~Adams, et~al.
\newblock Efficient propagation of systematic uncertainties from calibration to analysis with the {{SnowStorm}} method in {{IceCube}}.
\newblock \emph{Journal of Cosmology and Astroparticle Physics}, 2019\penalty0 (10):\penalty0 048, October 2019.
\newblock ISSN 1475-7516.
\newblock \doi{10.1088/1475-7516/2019/10/048}.

\bibitem[Aartsen et~al.(2017)Aartsen, Ackermann, Adams, Aguilar, Ahlers, Ahrens, Altmann, Andeen, Anderson, Ansseau, Anton, Archinger, Arg{\"u}elles, Auer, Auffenberg, Axani, Baccus, Bai, Barnet, Barwick, Baum, Bay, Beattie, Beatty, Tjus, Becker, Bendfelt, BenZvi, Berley, Bernardini, Bernhard, Besson, Binder, Bindig, Bissok, Blaufuss, Blot, Boersma, Bohm, B{\"o}rner, Bos, Bose, B{\"o}ser, Botner, Bouchta, Braun, Brayeur, Bretz, Bron, Burgman, Burreson, Carver, Casier, Cheung, Chirkin, Christov, Clark, Classen, Coenders, Collin, Conrad, Cowen, Cross, Day, Day, {de Andr{\'e}}, Clercq, Rosendo, Dembinski, Ridder, Descamps, Desiati, {de Vries}, {de Wasseige}, {de With}, DeYoung, {D{\'i}az-V{\'e}lez}, {di Lorenzo}, Dujmovic, Dumm, Dunkman, Eberhardt, Edwards, Ehrhardt, Eichmann, Eller, Euler, Evenson, Fahey, Fazely, Feintzeig, Felde, Filimonov, Finley, Flis, F{\"o}sig, Franckowiak, Fr{\`e}re, Friedman, Fuchs, Gaisser, Gallagher, Gerhardt, Ghorbani, Giang, Gladstone, Glauch, Glowacki, Gl{\"u}senkamp, Goldschmidt,
  Gonzalez, Grant, Griffith, Gustafsson, Haack, Hallgren, Halzen, Hansen, Hansmann, Hanson, Haugen, Hebecker, Heereman, Helbing, Hellauer, Heller, Hickford, Hignight, Hill, Hoffman, Hoffmann, Hoshina, Huang, Huber, Hulth, Hultqvist, In, Inaba, Ishihara, Jacobi, Jacobsen, Japaridze, Jeong, Jero, Jones, Jones, Joseph, Kang, Kappes, Karg, Karle, Katz, Kauer, Keivani, Kelley, Kemp, Kheirandish, Kim, Kim, Kintscher, Kiryluk, Kitamura, Kittler, Klein, Kleinfelder, Kleist, Kohnen, Koirala, Kolanoski, Konietz, K{\"o}pke, Kopper, Kopper, Koskinen, Kowalski, Krasberg, Krings, Kroll, Kr{\"u}ckl, Kr{\"u}ger, Kunnen, Kunwar, Kurahashi, Kuwabara, Labare, Laihem, Landsman, Lanfranchi, Larson, Lauber, Laundrie, Lennarz, Leich, {Lesiak-Bzdak}, Leuermann, Lu, Ludwig, L{\"u}nemann, Mackenzie, Madsen, Maggi, Mahn, Mancina, Mandelartz, Maruyama, Mase, Matis, Maunu, McNally, McParland, Meade, Meagher, Medici, Meier, Meli, Menne, Merino, Meures, Miarecki, Minor, Montaruli, Moulai, Murray, Nahnhauer, Naumann, Neer, Newcomb,
  Niederhausen, Nowicki, Nygren, Pollmann, Olivas, O'Murchadha, Palczewski, Pandya, Pankova, Patton, Peiffer, Penek, Pepper, de~los Heros, Pettersen, Pieloth, Pinat, Price, Przybylski, Quinnan, Raab, R{\"a}del, Rameez, Rawlins, Reimann, Relethford, Relich, Resconi, Rhode, Richman, Riedel, Robertson, Rongen, Roucelle, Rott, Ruhe, Ryckbosch, Rysewyk, Sabbatini, Herrera, Sandrock, Sandroos, Sandstrom, Sarkar, Satalecka, Schlunder, Schmidt, Schoenen, Sch{\"o}neberg, Schukraft, Schumacher, Seckel, Seunarine, Solarz, Soldin, Song, Spiczak, Spiering, Stanev, Stasik, Stettner, Steuer, Stezelberger, Stokstad, St{\"o}{\ss}l, Str{\"o}m, Strotjohann, Sulanke, Sullivan, Sutherland, Taavola, Taboada, Tatar, Tenholt, {Ter-Antonyan}, Terliuk, Te{\v s}i{\'c}, Thollander, Tilav, Toale, Tobin, Toscano, Tosi, Tselengidou, Turcati, Unger, Usner, Vandenbroucke, van Eijndhoven, Vanheule, van Rossem, van Santen, Vehring, Voge, Vogel, Vraeghe, Wahl, Walck, Wallace, Wallraff, Wandkowsky, Weaver, Weiss, Wendt, Westerhoff, Wharton,
  Whelan, Wickmann, Wiebe, Wiebusch, Wille, Williams, Wills, Wisniewski, Wolf, Wood, Woolsey, Woschnagg, Xu, Xu, Xu, Yanez, Yodh, Yoshida, and Zoll]{aartsenIceCubeNeutrinoObservatory2017}
M.G. Aartsen, M.~Ackermann, J.~Adams, et~al.
\newblock The {{IceCube Neutrino Observatory}}: Instrumentation and online systems.
\newblock \emph{Journal of Instrumentation}, 12\penalty0 (03):\penalty0 P03012, March 2017.
\newblock ISSN 1748-0221.
\newblock \doi{10.1088/1748-0221/12/03/P03012}.

\bibitem[{Adri{\'a}n-Mart{\'i}nez} et~al.(2016){Adri{\'a}n-Mart{\'i}nez}, Ageron, Aharonian, Aiello, Albert, Ameli, Anassontzis, Andre, Androulakis, Anghinolfi, Anton, Ardid, Avgitas, Barbarino, Barbarito, Baret, {Barrios-Mart{\'i}}, Belhorma, Belias, Berbee, Berg, Bertin, Beurthey, Beveren, Beverini, Biagi, Biagioni, Billault, Bond{\`i}, Bormuth, Bouhadef, Bourlis, Bourret, Boutonnet, Bouwhuis, Bozza, Bruijn, Brunner, Buis, Busto, Cacopardo, Caillat, Calamai, Calvo, Capone, Caramete, Cecchini, Celli, Champion, Moursli, Cherubini, Chiarusi, Circella, Classen, Cocimano, Coelho, Coleiro, Colonges, Coniglione, Cordelli, Cosquer, Coyle, Creusot, Cuttone, D'Amico, De~Bonis, De~Rosa, De~Sio, Di~Capua, Di~Palma, Garc{\'i}a, Distefano, Donzaud, Dornic, {Dorosti-Hasankiadeh}, Drakopoulou, Drouhin, Drury, Durocher, Eberl, Eichie, Eijk, Bojaddaini, Khayati, Elsaesser, Enzenh{\"o}fer, Fassi, Favali, Fermani, Ferrara, Filippidis, Frascadore, Fusco, Gal, Galat{\`a}, Garufi, Gay, Gebyehu, Giordano, Gizani, Gracia, Graf,
  Gr{\'e}goire, Grella, Habel, Hallmann, Haren, Harissopulos, Heid, Heijboer, Heine, Henry, {Hern{\'a}ndez-Rey}, Hevinga, Hofest{\"a}dt, Hugon, Illuminati, James, Jansweijer, Jongen, Jong, Kadler, Kalekin, Kappes, Katz, Keller, Kieft, Kie{\ss}ling, Koffeman, Kooijman, Kouchner, Kulikovskiy, Lahmann, Lamare, Leisos, Leonora, Clark, Liolios, Alvarez, Presti, L{\"o}hner, Lonardo, Lotze, Loucatos, Maccioni, Mannheim, Margiotta, Marinelli, Mari{\c s}, Markou, {Mart{\'i}nez-Mora}, Martini, Mele, Melis, Michael, Migliozzi, Migneco, Mijakowski, Miraglia, Mollo, Mongelli, Morganti, Moussa, Musico, Musumeci, Navas, Nicolau, Olcina, Olivetto, Orlando, Papaikonomou, Papaleo, P{\u a}v{\u a}la{\c s}, Peek, Pellegrino, Perrina, Pfutzner, Piattelli, Pikounis, Poma, Popa, Pradier, Pratolongo, P{\"u}hlhofer, Pulvirenti, Quinn, Racca, Raffaelli, Randazzo, Rapidis, Razis, Real, Resvanis, Reubelt, Riccobene, Rossi, Rovelli, Salda{\~n}a, Salvadori, Samtleben, Garc{\'i}a, Losa, Sanguineti, Santangelo, Santonocito, Sapienza,
  Schimmel, Schmelling, Sciacca, Sedita, Seitz, Sgura, Simeone, Siotis, Sipala, Spisso, Spurio, Stavropoulos, Steijger, Stellacci, Stransky, Taiuti, Tayalati, T{\'e}zier, Theraube, Thompson, Timmer, T{\"o}nnis, Trasatti, Trovato, Tsirigotis, Tzamarias, Tzamariudaki, Vallage, Elewyck, Vermeulen, Vicini, Viola, Vivolo, Volkert, Voulgaris, Wiggers, Wilms, Wolf, Zachariadou, Zornoza, and Z{\'u}{\~n}iga]{adrian-martinezLetterIntentKM3NeT2016}
S~{Adri{\'a}n-Mart{\'i}nez}, M~Ageron, F~Aharonian, et~al.
\newblock Letter of intent for {{KM3NeT}} 2.0.
\newblock \emph{Journal of Physics G: Nuclear and Particle Physics}, 43\penalty0 (8):\penalty0 084001, August 2016.
\newblock ISSN 0954-3899, 1361-6471.
\newblock \doi{10.1088/0954-3899/43/8/084001}.

\bibitem[Brown et~al.(2020)Brown, Mann, Ryder, Subbiah, Kaplan, Dhariwal, Neelakantan, Shyam, Sastry, Askell, Agarwal, {Herbert-Voss}, Krueger, Henighan, Child, Ramesh, Ziegler, Wu, Winter, Hesse, Chen, Sigler, Litwin, Gray, Chess, Clark, Berner, McCandlish, Radford, Sutskever, and Amodei]{brownLanguageModelsAre2020a}
Tom Brown, Benjamin Mann, Nick Ryder, et~al.
\newblock Language {{Models}} are {{Few-Shot Learners}}.
\newblock In \emph{Advances in {{Neural Information Processing Systems}}}, volume~33, pages 1877--1901. Curran Associates, Inc., 2020.

\bibitem[He et~al.(2022)He, Chen, Xie, Li, Dollar, and Girshick]{MAEVision}
Kaiming He, Xinlei Chen, Saining Xie, et~al.
\newblock { Masked Autoencoders Are Scalable Vision Learners }.
\newblock In \emph{2022 IEEE/CVF Conference on Computer Vision and Pattern Recognition (CVPR)}, pages 15979--15988, Los Alamitos, CA, USA, June 2022. IEEE Computer Society.
\newblock \doi{10.1109/CVPR52688.2022.01553}.
\newblock URL \url{https://doi.ieeecomputersociety.org/10.1109/CVPR52688.2022.01553}.

\bibitem[NVIDIA et~al.(2025)NVIDIA, Bjorck, Casta{\~n}eda, Cherniadev, Da, Ding, Fan, Fang, Fox, Hu, Huang, Jang, Jiang, Kautz, Kundalia, Lao, Li, Lin, Lin, Liu, Llontop, Magne, Mandlekar, Narayan, Nasiriany, Reed, Tan, Wang, Wang, Wang, Wang, Xiang, Xie, Xu, Xu, Ye, Yu, Zhang, Zhang, Zhao, Zheng, and Zhu]{nvidiaGR00TN1Open2025}
NVIDIA, Johan Bjorck, Fernando Casta{\~n}eda, et~al.
\newblock {{GR00T N1}}: {{An Open Foundation Model}} for {{Generalist Humanoid Robots}}, March 2025.

\bibitem[Timiryasov et~al.(2024)Timiryasov, Tastet, and Ruchayskiy]{timiryasovPolarBERTFoundationModela}
Inar Timiryasov, Jean-Loup Tastet, and Oleg Ruchayskiy.
\newblock {{PolarBERT}}: {{A Foundation Model}} for {{IceCube}}.
\newblock In \emph{Machine Learning and the Physical Sciences Workshop, NeurIPS}, 2024.
\newblock URL \url{https://ml4physicalsciences.github.io/2024/files/NeurIPS_ML4PS_2024_259.pdf}.

\bibitem[Yu et~al.(2026)Yu, Kamp, and Arg{\"u}elles]{yuReducingSimulationDependence2026a}
Felix~J. Yu, Nicholas Kamp, and Carlos~A. Arg{\"u}elles.
\newblock Reducing {{Simulation Dependence}} in {{Neutrino Telescopes}} with {{Masked Point Transformers}}, January 2026.

\bibitem[Pang et~al.(2022)Pang, Wang, Tay, Liu, Tian, and Yuan]{pangMaskedAutoencodersPoint2022}
Yatian Pang, Wenxiao Wang, Francis E.~H. Tay, et~al.
\newblock Masked {{Autoencoders}} for {{Point Cloud Self-supervised Learning}}, March 2022.

\bibitem[Bukhari et~al.(2023)Bukhari, Chakraborty, Eller, Ito, Shugaev, and {\O}rs{\o}e]{bukhariIceCubeNeutrinosDeep2023}
Habib Bukhari, Dipam Chakraborty, Philipp Eller, et~al.
\newblock {{IceCube}} -- {{Neutrinos}} in {{Deep Ice The Top}} 3 {{Solutions}} from the {{Public Kaggle Competition}}, October 2023.

\bibitem[{\O}rs{\o}e et~al.(2026){\O}rs{\o}e, {Meighen-Berger}, Lazar, Prado, {Moz{\'u}n-Mateo}, Rosted, Weigel, and Anaya]{orsoeNuBenchOpenBenchmark2026a}
Rasmus {\O}rs{\o}e, Stephan {Meighen-Berger}, Jeffrey Lazar, et~al.
\newblock {{NuBench}}: {{An}} open benchmark for deep learning-based event reconstruction in neutrino telescopes.
\newblock \emph{Journal of Instrumentation}, 21\penalty0 (05):\penalty0 T05001, May 2026.
\newblock ISSN 1748-0221.
\newblock \doi{10.1088/1748-0221/21/05/T05001}.

\bibitem[Kaplan et~al.(2020)Kaplan, McCandlish, Henighan, Brown, Chess, Child, Gray, Radford, Wu, and Amodei]{kaplanScalingLawsNeural2020}
Jared Kaplan, Sam McCandlish, Tom Henighan, et~al.
\newblock Scaling {{Laws}} for {{Neural Language Models}}.
\newblock https://arxiv.org/abs/2001.08361v1, January 2020.

\bibitem[Malyshkin(2023)]{malyshkinBaikalGVDNeutrinoTelescope2023}
Yury Malyshkin.
\newblock Baikal-{{GVD}} neutrino telescope: {{Design}} reference 2022.
\newblock \emph{Nuclear Instruments and Methods in Physics Research Section A: Accelerators, Spectrometers, Detectors and Associated Equipment}, 1050:\penalty0 168117, May 2023.
\newblock ISSN 0168-9002.
\newblock \doi{10.1016/j.nima.2023.168117}.

\bibitem[Mildenhall et~al.(2021)Mildenhall, Srinivasan, Tancik, Barron, Ramamoorthi, and Ng]{NeRF}
Ben Mildenhall, Pratul~P. Srinivasan, Matthew Tancik, et~al.
\newblock Nerf: representing scenes as neural radiance fields for view synthesis.
\newblock \emph{Commun. ACM}, 65\penalty0 (1):\penalty0 99--106, December 2021.
\newblock ISSN 0001-0782.
\newblock \doi{10.1145/3503250}.
\newblock URL \url{https://doi.org/10.1145/3503250}.

\bibitem[Paszke et~al.(2019)Paszke, Gross, Massa, Lerer, Bradbury, Chanan, Killeen, Lin, Gimelshein, Antiga, Desmaison, K{\"o}pf, Yang, DeVito, Raison, Tejani, Chilamkurthy, Steiner, Fang, Bai, and Chintala]{paszkePyTorchImperativeStyle2019}
Adam Paszke, Sam Gross, Francisco Massa, et~al.
\newblock {{PyTorch}}: {{An Imperative Style}}, {{High-Performance Deep Learning Library}}, December 2019.

\bibitem[Morales-Brotons et~al.(2024)Morales-Brotons, Vogels, and Hendrikx]{morales-brotons2024exponential}
Daniel Morales-Brotons, Thijs Vogels, and Hadrien Hendrikx.
\newblock Exponential moving average of weights in deep learning: Dynamics and benefits.
\newblock \emph{Transactions on Machine Learning Research}, 2024.
\newblock ISSN 2835-8856.
\newblock URL \url{https://openreview.net/forum?id=2M9CUnYnBA}.

\end{thebibliography}

\appendix

\section{A conservative lower bound on the number of pulses recorded per year}
\label{app:datavolume}

No collaboration publishes a pulse count for its trigger stream, so we bound it
from below using the trigger conditions themselves. Every trigger fixes a minimum
number of sensors that must register light before an event is written out.
Multiplying that minimum by the trigger rate and one year of livetime therefore
gives a very conservative floor on the number of pulses recorded per year.

\begin{center}
\small
\begin{tabular}{@{}llrr@{}}
\toprule
& \textbf{Trigger minimum} & \textbf{Rate} & \textbf{Pulses\,/\,year} \\
\midrule
IceCube~\citep{aartsenIceCubeNeutrinoObservatory2017}                        & $\geq 8$ \glspl{om} (SMT8) & $2.5$\,kHz & $6\times10^{11}$ \\
Baikal-GVD~\citep{malyshkinBaikalGVDNeutrinoTelescope2023} & $\geq 2$ adjacent \glspl{om} & $1.1$\,kHz & $7\times10^{10}$ \\
KM3NeT~\citep{adrian-martinezLetterIntentKM3NeT2016}   & $\geq 5$ L1 hits & $40$\,Hz & $6\times10^{9}$ \\
\midrule
Total & & & $\geq 7\times10^{11}$ \\
\bottomrule
\end{tabular}
\end{center}

Some $89\,\%$ of the total is IceCube's. The true figure is far larger, because the
bound credits each event with only the minimum number of sensors its trigger
demands, whereas a typical event lights many more, and with only a single pulse per
sensor, whereas a lit sensor often records several. It also counts only the three
detectors listed. The $7\times10^{11}$ is therefore a very conservative floor,
not an estimate.

\section{Architecture and training details}
\label{app:architecture}

This appendix documents the exact configuration behind every \texttt{CURTAIN} result in this
paper.

\begin{figure}[h]
  \centering
  \includegraphics[width=0.85\textwidth]{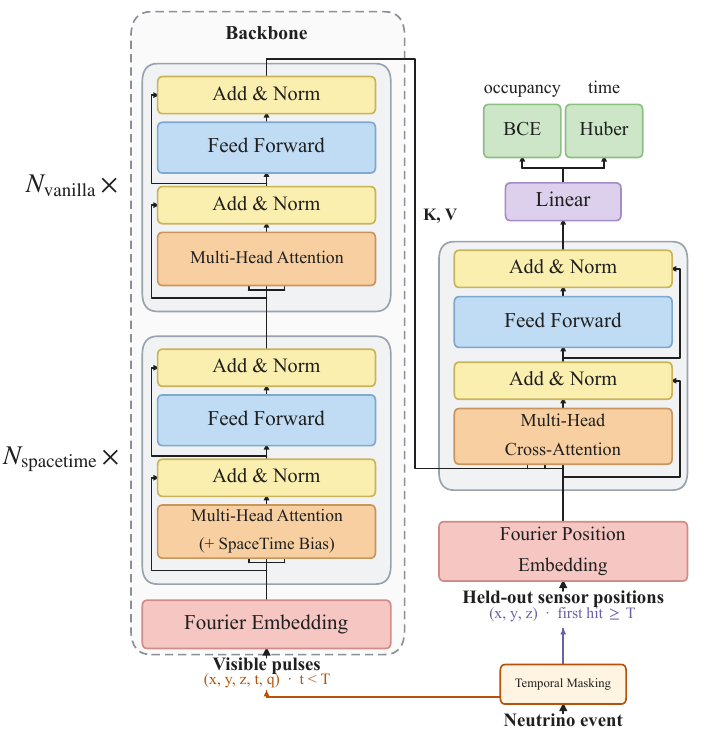}
  \caption{
    The \texttt{CURTAIN} pretraining architecture. Visible pulses enter the encoder
    backbone on the left; held-out sensor positions are embedded separately on
    the right and cross-attend to the encoded event, so the encoder never sees
    a mask token and the \texttt{SpaceTime} bias stays well defined.
  }
  \label{fig:architecture}
\end{figure}

\paragraph{Pulse cap.}
Events with more than $N_{\max}=768$ pulses are subsampled uniformly without
replacement to $N_{\max}$; smaller events are used in full. This bounds the
$\mathcal{O}(N^2)$ attention cost and is applied at every stage, with a fresh
subsample per capped event, and to all three encoders, so every comparison is
cap-matched. For \texttt{CURTAIN} the cap precedes the temporal split, so visible pulses
and queried sensors come from the same set. It is well above the median event of
137 pulses and binds on $22\,\%$ of test events, the high-energy tail (maximum
13{,}344).

\paragraph{Temporal masking.}
The cutoff $T$ is drawn so that a uniformly random fraction of the event's
pulses falls before it, re-drawn every epoch, so the model sees every masking
ratio equally often. Hard negatives are the nearest dark sensor to each
positive, with a smaller number drawn at random from the sensors that stay
dark.

\paragraph{\texttt{DeepIce} encoder.}
Visible pulses $(x,y,z,t,q)$, with times relative to the charge-weighted
mean time of the visible pulses, are embedded by a Fourier feature encoder
and processed by a reduced \texttt{DeepIce} encoder: $N_{\mathrm{spacetime}}=2$
multi-head attention blocks whose attention logits carry an additive
relative space--time interval bias (\texttt{SpaceTime} bias), followed by
$N_{\mathrm{vanilla}}=3$ standard pre-norm Transformer blocks
(multi-head attention, feed-forward, residual connections and
normalization). We use $d_{\mathrm{model}}=128$ with $8$ heads of width
$16$, for $2.2$\,M parameters. The backbone outputs one token per visible
pulse. See Fig. \ref{fig:architecture} for a sketch of the architecture.

\paragraph{Query head (decoder).}
Held-out sensor positions are lifted by a NeRF-style Fourier position
embedding~\citep{NeRF} and attend to the encoded event through a single cross-attention
block: queries come from the embedded positions, keys and values from the
per-pulse tokens, followed by a feed-forward layer and a linear projection
to the per-query outputs. The head adds $0.23$\,M parameters, for $2.5$\,M
during pretraining, and is discarded afterwards, so only the $2.2$\,M
encoder transfers downstream.
\paragraph{Pretraining objectives.}
Each query is scored by two objectives,
\begin{equation}
\mathcal{L} \;=\; \mathcal{L}_{\mathrm{occ}}
\;+\; \lambda_{t}\,\mathcal{L}_{\Delta t},
\end{equation}
where $\mathcal{L}_{\mathrm{occ}}$ is a \gls{bce} on whether the queried
sensor lights up after $T$, and $\mathcal{L}_{\Delta t}$ is a Huber loss on
the first-hit time of hit sensors. The time term applies to hit queries only and $\lambda_{t}$ is chosen as 1. 

\paragraph{Optimization.} We train with Adam ($\beta_1=0.9$, $\beta_2=0.999$, no weight decay) at a learning rate of $1.4\times10^{-3}$ for a batch size of 256 events, with no warmup; the learning rate is halved whenever the held-out \gls{ssl} loss has not improved for 10 epochs. Training runs for at most 200 epochs, with early stopping on the same loss at patience 15. Gradients are clipped at a global norm of 1.0 and we train in full \texttt{float32} on one to four NVIDIA H200s, taking 330 GPU hours for the largest pretraining set (3.9 million events).

\paragraph{Baselines.} The \texttt{Neptune} and \texttt{PolarBERT} baselines use the same optimizer, schedule and budget.

\section{Finetuning details}\label{app:finetuning}

For downstream tasks the query head is removed and the backbone is
finetuned end-to-end with a fresh task head, using the same input convention as pretraining. For the \texttt{DeepIce} encoders, pretrained and from scratch alike, the event embedding is the mean of the encoder's output pulse tokens; \texttt{Neptune} and \texttt{PolarBERT} keep the readout of their original implementations. The task head is a layer normalization followed by a single linear layer, computed in \texttt{float32}. The energy task head predicts the logarithmic neutrino energy. The direction task head predicts four values: the three Cartesian components of the neutrino direction, normalized to a unit vector, and $\log_{10}(\kappa -1)$, where $\kappa$ is the concentration parameter of the \gls{vmf} distribution.

For the pretrained encoders, the backbone is frozen for the first two epochs so that only the new task head is trained; the from-scratch \texttt{DeepIce} is trained end-to-end from the start. All finetuning runs use a batch size of 256 and a learning rate of $5\times10^{-4}$ without weight decay, reduced on plateau (\texttt{ReduceLROnPlateau} from \texttt{PyTorch}~\citep{paszkePyTorchImperativeStyle2019}), and an exponential moving average of the weights~\citep{morales-brotons2024exponential} with a decay rate of 0.999. We validate every 300 steps and stop once the validation loss has not improved for 8 validations, or after at most 100 thousand steps. The cross-geometry finetune of the encoder pretrained on 3.9 million events additionally uses a 2{,}000-step learning-rate warm-up.

\section{Pretraining schemes compared}
\label{app:schemes}

All three schemes hide part of an event and train the model to recover what was
hidden, but they differ along four axes: which pulses are hidden, which of their
information is hidden, what the model must then predict, and which architectures
the scheme admits. Table~\ref{tab:masking} summarizes them.

\paragraph{\texttt{PolarBERT}.}
A random $25\,\%$ of pulses are masked. Each pulse is represented by a concatenation of a token identifying the \gls{om} that recorded it and a \gls{mlp} on the pulse features. \texttt{PolarBERT} hides the \gls{om} identifying token. The model then recovers it by cross-entropy over the vocabulary of \glspl{om}, with an auxiliary regression of the total charge. Because the vocabulary enumerates the sensors of one detector, carrying the pretrained model to another geometry means learning a new vocabulary.

\paragraph{\texttt{Neptune}.}
What is masked depends on the downstream task. For track direction the arrival
time of every pulse is hidden; for cascade classification a random $75\,\%$ of
pulses have both time and position hidden. In either case the model regresses the
masked coordinates with a smooth-L1 loss. Our baseline uses the track-direction
variant, with the arrival time of every pulse hidden. The scheme has no per-detector
vocabulary and so carries over to another geometry in principle, but
cross-detector transfer has not been reported.

\paragraph{\texttt{CURTAIN}.}
Every pulse after a sampled threshold time is hidden and all of its information
discarded. Rather than reconstruct the hidden pulses, the model is queried at
sensor positions and must separate \glspl{om} that light up after the cut from
\glspl{om} that stay dark for the whole event. A query carries only a position,
so the scheme depends on no sensor numbering and transfers to a detector it was
not pretrained on (Figure~\ref{fig:finetune_scaling}, right).

\paragraph{Architecture.}
\texttt{PolarBERT} and \texttt{Neptune} mask pulses in place and use encoder-only transformers, so
the masked pulses pass through the encoder's attention. \texttt{CURTAIN}'s encoder sees
only the visible pulses and its decoder is queried separately, leaving the
encoder free of mask tokens and free to use an attention bias defined on
space--time distance.

\begin{table}[!t]
  \centering
  \small
  \setlength{\tabcolsep}{6pt}
  \renewcommand{\arraystretch}{1.15}
  \caption{\glsentryshort{ssl} pretraining schemes for neutrino telescopes.
    \texttt{Neptune}'s slashed entries read as track direction / cascade
    classification; our baseline uses the track-direction variant.}
  \label{tab:masking}
  \begin{tabularx}{\textwidth}{@{}l *{3}{>{\raggedright\arraybackslash}X}@{}}
    \toprule
      & \textbf{\texttt{PolarBERT}} & \textbf{\texttt{Neptune}} & \textbf{\texttt{CURTAIN}} \\
    \midrule
    Which pulses
      & Random 25\,\% of pulses
      & All / random 75\,\%
      & Every pulse after a threshold time $T$ \\
    Which information
      & The \acrshort{om} token only
      & Time / time and space
      & All pulse information \\
    How reconstructed
      & Cross-entropy over the \acrshort{om} vocabulary, plus an auxiliary total-charge regression
      & Smooth-L1 regression of the masked coordinates
      & Binary discrimination of active \glsentryshortpl{om} against sampled negatives, plus an auxiliary first-hit-time regression \\
    Transformer type
      & Encoder only
      & Encoder only
      & Encoder--decoder \\
    \midrule
    Cross-detector
      & New vocabulary required
      & Untested
      & Demonstrated \\
    \bottomrule
  \end{tabularx}
\end{table}

\end{document}